\documentclass[a4paper,12pt]{article}
\usepackage{soul}
\usepackage[usenames,dvipsnames]{color}
\usepackage{jheppub}
\usepackage{esvect}
\usepackage{amsmath, amssymb, slashed, epsf, color, graphicx, latexsym, tensor, physics, xcolor}
\usepackage{epsfig}
\usepackage{comment}
\usepackage{graphics}
\usepackage{mathtools}
\usepackage{appendix}
\usepackage{inputenc}
\usepackage{indentfirst}
\usepackage{kbordermatrix}
\usepackage{booktabs}
\def\nn{\nonumber}

\title{\boldmath Logarithmic corrections to entropy of 3D cosmological solutions from celestial dual}
\author[1]{Arindam Bhattacharjee,}
\author[2]{Shruti Menon,}
\author[2,3]{Muktajyoti Saha}

\affiliation[1]{Faculty of Physics, University of Warsaw, \\
ul. Pasteura 5, 02-093 Warsaw, Poland}
\affiliation[2]{Indian Institute of Science Education and Research Bhopal,\\ Bhopal Bypass, Bhauri, Bhopal 462066, India}
\affiliation[3]{International Centre for Theoretical Sciences (ICTS-TIFR),\\
Tata Institute of Fundamental Research, Shivakote, Bengaluru 560089, India}

\emailAdd{arindamb.hep@gmail.com}
\emailAdd{shruti98.menon@gmail.com}
\emailAdd{muktajyoti.hepth@gmail.com}

\abstract{Recently a one-dimensional Schwarzian type theory was proposed as an effective dual theory of pure gravity in (2+1) dimensional asymptotically flat spacetimes \cite{Bhattacharjee:2023sfd}. This codimension-two `celestial' dual captures the Bekenstein-Hawking entropy of bulk flat cosmologies in semiclassical limit. In this paper, we extend this analysis beyond semiclassical approximation and evaluate the one-loop exact partition function of this celestial dual theory. Our
analysis results in novel nontrivial logarithmic corrections to the area term of entropy, appearing from the one-loop path integral.}

\begin{document} 
\maketitle

\section{Introduction}

Lower dimensional gravitational theories, with or without cosmological constant, serve as useful playgrounds to understand various properties of gravity in our own universe. These theories are usually much simpler but contains a lot of the features of higher dimensional systems. Pure gravity in (2+1) dimensions, for example, has no local gravitons but have rich asymptotic structures that have forwarded the understanding of holography in AdS and non-AdS spacetimes \cite{Brown:1986nw, Strominger:1997eq, Barnich:2006av}. In this work, we focus on asymptotically flat spacetimes in (2+1) dimensions. The physics of these spacetimes and their dual descriptions has been extensively studied  in terms of two dimensional field theories \cite{Bagchi:2009my, Bagchi:2009pe, Barnich:2012xq, Bagchi:2012xr, Barnich:2012rz, Barnich:2015sca, Banerjee:2019lrv, Barnich:2017jgw, Merbis:2019wgk}, closely resembling the usual AdS$_3$/CFT$_2$ correspondence. We focus on a \textit{celestial} holographic nature of (2+1)-dimensional gravity with zero cosmological constant, where the dual field theory lives on a 1D celestial circle rather than a two dimensional surface i.e. the null infinity. The idea of celestial holography generally is that a theory of gravity in $(d+1)$-dimensional asymptotically flat spacetime is dual to a conformal field theory in $(d-1)$-dimensional celestial sphere or the spatial cross-section of null infinity. There has been a lot of progress in understanding this duality in (3+1)-dimensions by studying 4D scattering amplitudes in terms of a celestial CFT$_2$ \cite{deBoer:2003vf,
Pasterski:2016qvg, Pasterski:2017kqt,
Strominger:2013jfa, He:2014laa, Kapec:2014opa,
Cheung:2016iub, Kapec:2016jld}. Our motivation is to gather insights into the Celestial nature of the holographic dual of flat spacetimes by extending the idea to (2+1) dimensions.

The phase space of (2+1)D pure gravity with zero cosmological constant is spanned by boundary modes corresponding to its asymptotic symmetries. Under Brown-Henneaux type boundary conditions, these symmetries form the BMS$_3$ group \cite{Barnich:2010eb} which consists of supertranslations and superrotations analogous to (3+1) dimensions. Pure superrotations form a Virasoro subalgebra and pure supertranslations form an abelian subalgebra. This theory does not contain black holes \cite{Ida:2000jh} but there are non trivial saddle points other than the usual Minkowski space. These are called Flat Space Cosmologies (FSC) that have cosmological horizons. These FSC solutions are labeled by mass and angular momentum parameters. They mimic the thermodynamic properties of black holes \cite{Barnich:2012aw, Barnich:2012xq}.

As is typical for asymptotic symmetries, BMS$_3$ symmetries can be realized as asymptotic Killing vectors of the bulk and diffeomorphisms of the null boundary. The key property is that the transformations only nontrivially depend on the coordinate of the celestial circle. Thus BMS$_3$ group can be purely described by vector fields living on the celestial circle \cite{Barnich:2014kra, Barnich:2015uva, Oblak:2016eij}. Exploiting this structure, some of the authors of this manuscript have shown that the phase space of 3D gravity with zero cosmological constant near an FSC saddle in the large charge regime, can have an emergent celestial field theoretic description \cite{Bhattacharjee:2023sfd}. This Schwarzian-like theory is the dynamical theory of pseudo-Goldstone modes associated with the BMS$_3$ group\footnote{This pattern can be seen in \cite{Turiaci:2016cvo, Halyo:2019zek} where the holomorphic part of a CFT$_2$ was described in terms of a Schwarzian theory of Pseudo Goldstone modes of Virasoro symmetry.}. The Schwarzian part of the theory appears from the superrotations that act as diffeomorphisms on the circle. This part was also obtained from an independent analysis of holographic reduction\footnote{This reduction crucially involves the ideas of wedge holography \cite{akal_codimension_2020, geng_jackiw-teitelboim_2022}.} of bulk superrotated spacetimes \cite{Bhattacharjee:2022pcb}. The supertranslation sector results in an additional Lagrange multiplier field in the theory. Recently it was shown that 1D Schwarzian type actions also emerge in flat space fluid/gravity correspondence in 3D \cite{Adami:2024rkr} (See \cite{Garcia-Sepulveda:2022lga}, for a 1D celestial description of 3D QFTs).

It was shown in \cite{Bhattacharjee:2023sfd} that the thermodynamic properties \cite{Barnich:2012xq, Bagchi:2012xr} of FSC solutions with large charges can be correctly obtained from a semiclassical analysis of the path integral of the celestial dual field theory. The matching of Bekenstein-Hawking entropy strongly supports the validity of a celestial dual. Both the area law of entropy and its logarithmic corrections are universal features of gravitational theories \cite{Solodukhin:1994yz,
Kaul:2000kf,
Medved:2004eh, Banerjee:2010qc,Sen:2012dw}. Thus the computation of these contributions in entropy is a significant problem in understanding holographic dualities. The main goal of this paper is to extend the analysis of \cite{Bhattacharjee:2023sfd} in finding the logarithmic corrections to the entropy of FSC solutions from celestial side. For this, we would like to compute the path integral of the celestial theory. The Lagrange multiplier field in the celestial dual theory makes the path integral one-loop exact. We compute the one-loop path integral by coadjoint orbit quantization, following the quantization of the Schwarzian theory in the context of Virasoro group \cite{Stanford:2017thb}. This was generalized to other infinite dimensional groups e.g. warped Schwarzian theory corresponding to Warped Virasoro \cite{Afshar:2019tvp} and bms$_2$-Schwarzian theory corresponding to the BMS$_2$ symmetry of flat JT gravity \cite{Afshar:2021qvi}. 

Our analysis results in novel logarithmic corrections to the entropy of FSC solutions, which come from the one-loop partition function\footnote{Log corrections to microcanonical entropy can appear from the one-loop partition function and also from change of ensemble while transforming the partition function to density of states. The latter being a consequence of the choice of ensemble, is in some sense trivial. Whereas the former contributes in all choices of ensembles. This is elaborated in Sec. \ref{sec-ent}}. The coefficient of the log correction differs from earlier results \cite{Bagchi:2013qva} because the log corrections from the partition function were not taken into account in those computations. 

The rest of the paper is organized as follows: In section \ref{sec-pgb-act}, we summarize the details of the 1D celestial theory. In section \ref{sec-1loop}, we compute the path integral of this theory. In section \ref{sec-ent}, we compute the logarithmic corrections to microcanonical entropy. We discuss the results in section \ref{sec-concl} and briefly compare the relations with relevant computations for BTZ black holes. Appendix \ref{zeta-reg} discusses the regularization of an infinite product that is important for the main computations.

\section{Celestial description of 3D gravity near FSC saddle}\label{sec-pgb-act}

\subsection{Geometry and Thermodynamics of FSC}
The stage for our analysis are flat space cosmologies (FSC) which are non-trivial vacuum solutions of Einstein equations in 3D. These were first described in \cite{Barnich:2012aw} as a suitable flat space limit of BTZ black holes in AdS$_3$ spacetime. The general metric is in ADM form,
\begin{align}\label{fsc-adm}
ds^2 = -N^2 dt^2 + \frac{1}{N^2} dr^2 +r^2 (d\theta + N^{\theta} dt)^2
\end{align}
with $N^2 = -8MG_N + \frac{16G_N^2 J^2}{r^2}$ and $N^{\theta} = \frac{4G_N J}{r^2}$. $M,J$ are the mass and angular momentum of the spacetime. Global Minkowski space is also part of this family of metric with $M = -1/8G_N$ and $J=0$ but it is isolated from the general FSC solutions with $M>0$. The general solution has a horizon at $r = \sqrt{\frac{2G_N J^2}{M}}$. Outside this horizon, the spacetime is 'cosmological', i.e. the radius of the spatial section expands in time.

We may use the standard procedures to study the thermodynamic properties of this horizon. It is a Killing Horizon and the Hawking temperature of the horizon is $T_H = \frac{2}{\pi} \sqrt{\frac{2G_N M^3}{J^2}}$ and the Bekenstein Hawking entropy is given by, $S = \frac{\pi |J|}{\sqrt{2G_NM}}$. The angular velocity at the horizon is given by $\Omega = -\frac{2M}{J}$. This is the extent of semiclassical thermodynamics. For general gravitational systems, the next to leading order term in entropy of a horizon is proportional to the logarithm of area. The coefficient of this term is in general fixed by the infrared physics of the governing theory \cite{Sen:2012dw}. The goal of our work is to write a dual description of the gravitational system that faithfully captures both the semiclassical thermodynamics and also the logarithmic corrections.

\subsection{Dual action and Semiclassical entropy matching}
We begin by summarizing the celestial dual description obtained in \cite{Bhattacharjee:2023sfd}. By this we mean an effective description of the dynamics of pure gravity in the (2+1) dimensional asymptotically flat bulk spacetime, via a theory living on the Celestial circle. The realisation of BMS$_3$ group on a circle is already very well studied
\cite{Barnich:2014kra, Barnich:2015uva, Oblak:2016eij} and these notions form the basis of our dual construction. From this perspective, the superrotation part of BMS$_3$ generate diffeomorphisms of the circle forming the Diff($S^1$) or Virasoro subgroup. The supertranslations are abelian vector fields on the circle. Each element of the group is given by $(f,\lambda;\alpha,\mu)$ where $f(\theta)$ and $\alpha(\theta)$ parameterise superrotation and supertranslation respectively. Their corresponding central elements are given by $\lambda$ and $\mu$. The infinitesimal transformations form the $\mathfrak{bms}_3$ algebra and it is generated by vector fields on the circle.

We can define a dual vector space (coadjoint space) with elements $(j,c_1;p,c_2)$ where $j$ is the stress tensor dual to infinitesimal diffeomorphisms and $p$ is the conserved current corresponding to supertranslations. Their modes satisfy the commutation relations,
\begin{align}
    &[j_m,j_n] = (m-n)j_{m+n} + \frac{c_1}{12}m^3\delta_{m+n,0} 
 \nonumber\\
    & [j_m,p_n] = (m-n)p_{m+n} + \frac{c_2}{12}m^3\delta_{m+n,0} \nonumber \\
    & [p_m,p_n] = 0. \label{bms3-algebra} 
\end{align}
The above commutation relations also highlight the semidirect product structure of $\mathfrak{bms}_3$ algebra. $c_1 = 0, c_2 = \frac{3}{G_N}$ corresponds to pure gravity\footnote{$c_1 \neq 0$ corresponds to adding a Lorentz Chern-Simons term with the gravity action.}.\\

Now we consider a representation of this algebra. The states are labeled by quantum numbers $\ket{h_j,h_p}$ which are the eigenvalues of the mutually commuting zero modes $j_0$ and $p_0$. This state represents a particular Flat cosmology in the bulk with mass $M=h_p$ and angular momentum $J=h_j$.

In the large charge regime ${h_j,h_p\gg c_1, c_2}$, the state has a large degeneracy. Hence, this can be thought of as an ensemble labeled by a finite temperature $\beta$ and a chemical potential $\Phi$, which arise due to the stress tensor and the weight-2 current respectively. As mentioned earlier and explicitly shown in \cite{Bhattacharjee:2023sfd}, BMS transformations can be interpreted as diffeomorphisms and gauge transformations on the celestial circle. It turns out that the periodicities of the celestial circle and the corresponding gauge fiber are related to the parameters $\Phi$ and $\beta$, following the thermodynamic relations:
\begin{align}
         h_p = \frac{c_2\pi^2}{6\Phi^2}, \quad h_j = \frac{c_1\pi^2}{6\Phi^2} - \frac{c_2\pi^2}{3}\frac{\beta}{\Phi^3}.  \label{parameters-relation}
\end{align}

Since the state $\ket{h_j,h_p}$ correspond to a FSC solution in the bulk, the BMS symmetry is spontaneously broken to the global symmetry of the solutions. BMS symmetry is also anomalously broken here due to nonzero central charges. Thus the dynamics in this part of the Hilbert space is expected to be described by the dynamics of pseudo-Goldstone modes corresponding to softly broken BMS$_3$ symmetry. The Goldstone modes are in one-to-one correspondence with BMS$_3$ transformation parameters $f$ and $\alpha$ which are promoted to local quantum fields (following \cite{Turiaci:2016cvo}). The coadjoint action of the group then expresses the conserved currents of this theory in terms of these quantum fields. From this, the effective action was written in \cite{Bhattacharjee:2023sfd} as,
\begin{align}
       I = &-\frac{\beta}{2\pi}\int_0^{2\pi} d\theta\left([f'(\theta)]^2 h_p - \frac{c_2}{12}S[f](\theta) \right) \nonumber \\
       &-\frac{\Phi}{2\pi}\int_0^{2\pi} d\theta \Big[(f'(\theta))^2 h_j - \frac{c_1}{12}S[f](\theta) - \gamma(\theta)\left( 2f''(\theta) h_p - \frac{c_2}{12}\frac{S[f]'(\theta)}{f'(\theta)}\right)\Big] \label{bms-PGB}
\end{align}
where $\gamma \equiv \alpha\circ f$ acts as a Lagrange multiplier. The parameters $h_j$ and $h_p$ are related to $\beta$ and $\Phi$ via \eqref{parameters-relation}. The Grand cannonical partition function of the theory can be computed as a path integral of this action. It turns out that under semiclassical approximation, the solution $f(\theta) = \theta$ of this theory computes the correct thermodynamics of bulk FSC solutions. The Bekenstein-Hawking entropy is given as,
\begin{align}
       S_{\text{BH}} = \frac{\pi}{\sqrt{6}}\left(\sqrt{\frac{c_2}{h_p}}h_j + c_1\sqrt{\frac{h_p}{c_2}}\right). \label{BH-ent}
\end{align}
which for $c_1 = 0$, exactly boils down to the FSC entropy we discussed above. Now we will consider the full path integral of the theory \eqref{bms-PGB}.

\section{Quantization of celestial dual theory}\label{sec-1loop}
As discussed in the last section, the BMS$_3$ symmetry is softly broken by the finite temperature and chemical potential and it is non linearly realised by the pseudo-Goldstone modes. For the global Minkowski vacuum, which is represented by the state $\ket{h_j = 0, h_p = -1/8G_N}$, the symmetry is broken to Poincar\'{e} group, but that is not preserved for general FSC states. However, some part of the global Poincar\'{e} symmetry should still be preserved corresponding to the symmetries of the saddle point. In the dual theory side, this preserved part acts as a gauge symmetry of the dynamical theory of pseudo-Goldstone modes. Let us first discuss these symmetries.

\subsection{Gauge symmetries of the theory}
To understand the symmetries of the state $(h_j,h_p)$, we consider the coadjoint action of the BMS$_3$ group \cite{Oblak:2016eij},
\begin{align}
    p(\theta) &= [f'(\theta)]^2\Tilde{p}(f(\theta)) - \frac{c_2}{12}S[f](\theta), \label{p-trans}\\
    j(\theta) &= [f'(\theta)]^2\Tilde{j}(f(\theta)) - \frac{c_1}{12}S[f](\theta) \nonumber \\
    & + f'(\theta) \left[\alpha\circ f (\tilde{p}\circ f)' + 2(\alpha\circ f)' \tilde{p}\circ f \right](\theta)  -\frac{c_2}{12}[f'(\theta)]^2 \alpha'''\circ f(\theta). \label{j-trans}
\end{align} 
Here, $S[f]$ denotes the Schwarzian derivative of $f$. The coadjoint action of the $\mathfrak{bms}_3$ algebra is given by the infinitesimal version of the above relations. For this we take the BMS transformation parameters as $f \xrightarrow[]{\text{inf.}} 1 + \epsilon (\theta)$ and $\alpha \xrightarrow[]{\text{inf.}} \sigma (\theta)$. With this the transformation becomes,
\begin{align}
    & \delta j = \epsilon j' + 2\epsilon' j - \frac{c_1}{12}\epsilon''' + \sigma p' + 2\sigma' p - \frac{c_2}{12}\sigma''' \\
    & \delta p = \epsilon p' + 2\epsilon' p - \frac{c_2}{12}\epsilon'''
\end{align}
 Our states are labelled by the zero modes of the $(j(\theta),p(\theta))$ modes, so we are interested in the global symmetries of constant modes. For symmetry transformation, the above variations should vanish. Thus we consider the transformation of arbitrary constant representatives $(h_j,h_p)$ and set them to zero:
\begin{align}
     \delta p &= 2\epsilon' h_p - \frac{c_2}{12}\epsilon''' = 0 
\end{align}
The solution to this is given by,
\begin{align}\label{symmetry-epsilon}
     \epsilon (\theta) &= k_1 \text{e}^{\sqrt{\frac{24h_p}{c_2}}\theta} + k_2 \text{e}^{\sqrt{\frac{24h_p}{c_2}}\theta} + k_3
\end{align}
Hence we see there are three independent solutions in general. But we must recall that since $\epsilon (\theta)$ parameterises the diffeomorphism along the circle, it must be periodic with a period of $2\pi$. The periodicity condition on $\epsilon$ fixes $h_p = -\frac{n^2 c_2}{24}$ for integer $n$. We see that the invariance under the global superrotations i.e. the $n = 0, \pm 1$ modes of $\epsilon$ further fixes $h_p = -\frac{c_2}{24}$. Now we demand similar invariance of $h_j$ to further achieve invariance under global supertranslations.  
\begin{align}
    \delta j & = 2\epsilon' h_j - \frac{c_1}{12}\epsilon''' + 2\sigma' h_p - \frac{c_2}{12}\sigma''' \nonumber \\
    & = \frac{1}{12}\left[i(c_1 + 24h_j)(k_1\text{e}^{i\theta} - k_2\text{e}^{-i\theta}) - c_2\sigma'-c_2\sigma'''\right] 
\end{align}
Solving $\delta j = 0$, we get
\begin{align}\label{symmetry-sigma}
    \sigma &= \frac{c_1 + 24h_j}{4c_2} [k_1\text{e}^{i\theta}(3-2 i\theta ) + k_2\text{e}^{-i\theta}(3+2i\theta)] + l_1\text{e}^{i\theta} + l_2\text{e}^{-i\theta} + l_3
\end{align}
The periodicity condition on $\sigma$ fixes $h_j = -\frac{c_1}{24}$ and this automatically fixes $\sigma$ as a linear combination of the global supertranslations. Thus, the full Poincar\'{e}-invariant representative is,
\begin{align}
    h_j = -\frac{c_1}{24}; \quad h_p = -\frac{c_2}{24} \label{poinc inv rep}
\end{align}
These values correspond to the Minkowski vacuum. We can easily see that for arbitrary constants $(h_j,h_p)$, only the $u(1)\oplus u(1)$ subalgebra of the global Poincar\'{e} algebra is preserved which corresponds to the constant pieces in (\ref{symmetry-epsilon}) and (\ref{symmetry-sigma}). In the bulk side, these correspond to time translation ($\partial_t$) and axial rotation ($\partial_{\theta}$) symmetries of FSC metric (\ref{fsc-adm}).

\subsection{Path integral measure}
The state corresponding to FSC saddle is invariant under the $U(1)\times U(1)$ subgroup of the BMS$_3$ symmetry group. The 1D effective 'celestial' theory of the pseudo-Goldstone modes around this saddle inherits this symmetry as a gauge symmetry of the theory. So, the modes $(f,\alpha)$ has this gauge symmetry and when we perform the path integral over all possible field configurations, only the physically distinct configurations should contribute. This implies we need to quotient all possible configuration with the $U(1)\times U(1)$ symmetry and this must be the stabilizer of the orbit.


Firstly, we would like to compute the path integral measure. It is well known that coadjoint orbits are symplectic manifolds and they have a natural notion of measure derived from the symplectic structure. The technique has been used in the quantization of various Schwarzian theories \cite{Stanford:2017thb,Afshar:2019tvp,Afshar:2021qvi}. Secondly, our celestial dual theory \eqref{bms-PGB} involves a Lagrange multiplier field $\gamma = \alpha\circ f$, which simplifies the computation. This field trivially enforces one-loop exactness\footnote{This feature is also present in the BMS$_2$-Schwarzian \cite{Afshar:2021qvi}.}. Hence, we will find the integration measure for small fluctuation around the classical saddle $f(\theta) = \theta, \gamma(\theta) = 0$ of the theory \eqref{bms-PGB}.

Our state $\ket{h_j,h_p}$ represents a particular orbit in the coadjoint space which is given by $(h_j,c_1;h_p,c_2)$. The other members of this orbit can be parameterised by BMS$_3$ transformations $(f,\alpha)$ acting on the state. We obtain this parametrization from the coadjoint action \eqref{p-trans} and \eqref{j-trans},
\begin{align}
     p(\theta) &= h_p [f'(\theta)]^2 -\frac{c_2}{12}S[f](\theta) ,\nn\\
    j(\theta) &= h_j [f'(\theta)]^2 -\frac{c_2}{12}S[f](\theta)\nn\\
    & + \bigg[h_p f'\gamma' - \frac{c_2}{12} (f')^2\left( \frac{\gamma'''}{f'} - \frac{3f''}{f'^2}\gamma'' + \frac{3f''^2}{f'^3}\gamma' - \frac{f'''}{f'^2}\gamma' \right)\bigg](\theta)
\end{align}
where the general element is of the form $(j,c_1;p,c_2)$. We have replaced $\alpha$ in terms of $\gamma$ as this is the basic field that appears in the action. Now, we will find the symplectic structure on this coadjoint orbit. For this, we consider the standard Kirillov-Kostant-Souriau symplectic form. At any point $q$ on the coadjoint orbit, this form is given by,
\begin{equation}
    \omega_q(ad^*_Xq,ad^*_Yq)=\langle q,[X,Y]\rangle, \label{KK-def}
\end{equation}
where $ad*_X q, ad*_Y q$ represents elements of the tangent space at the point q labelled by the corresponding Lie algebra elements $X,Y$. $[,]$ is the binary operation on the Lie algebra. $\langle\cdot,\cdot\rangle$ denotes the inner product between adjoint and coadjoint elements.

The bracket on the $\mathfrak{bms}_{3}$ Lie algebra is given as,
\begin{align}
    & [X_1,X_2] = ([\epsilon_1,\epsilon_2], c(\epsilon_1,\epsilon_2); [\epsilon_1,\sigma_2] - [\epsilon_2,\sigma_1], c(\epsilon_1,\sigma_2) - c(\epsilon_2,\sigma_1)) \\
    & c(\epsilon_1,\epsilon_2) \equiv -\frac{1}{12}\langle s[\epsilon_1],\epsilon_2 \rangle  = -\frac{1}{24\pi}\int_{S^1}\epsilon_1\epsilon'''_2
\end{align}

Here $X\equiv (\epsilon,a;\sigma,b)$ labels the elements of the Lie algebra. The inner product between coadjoint and adjoint spaces are given as,
\begin{align}
    \langle(j,c_1;p,c_2), (\epsilon,a;\sigma,b)\rangle = \frac{1}{2\pi}\int_{S^1}(j\epsilon + p\sigma) + a c_1 + b c_2
\end{align}
Thus using \eqref{KK-def}, we write down the component of the symplectic form on $\mathcal{O}_{(h_j,h_p)}$ along the Hamiltonian vector fields corresponding to the Lie algebra elements $X_1$ and $X_2$:
\begin{align}
    \omega_{12} =& -\frac{c_1}{24\pi}\int_{S^1} \left[S[f](\epsilon_1\epsilon'_2 - \epsilon_2\epsilon'_1) - \epsilon'_1\epsilon''_2\right] -\frac{c_2}{24\pi}\int_{S^1} \Big[ S[f](\epsilon_1\sigma'_2 - \sigma_2\epsilon'_1 - \epsilon_2\sigma'_1 + \sigma_1\epsilon'_2) \nonumber \\
    & + \left(\frac{\gamma'''}{f'} - \frac{3f''}{f'^2}\gamma'' + \frac{3f''^2}{f'^3}\gamma' - \frac{f'''}{f'^2}\gamma' \right)(\epsilon_1\epsilon'_2 - \epsilon_2\epsilon'_1) - (\epsilon'_1\sigma''_2 - \epsilon'_2\sigma''_1)\Big] \nn\\
    & +  \frac{h_p}{2\pi}\int_{S^1} \left[f'^2(\epsilon_1\sigma'_2 - \sigma_2\epsilon'_1 - \epsilon_2\sigma'_1 + \sigma_1\epsilon'_2) +f'\gamma' (\epsilon_1\epsilon'_2 - \epsilon_2\epsilon'_1)\right]\nn\\
    & +\frac{h_j}{2\pi}\int_{S^1} f'^2(\epsilon_1\epsilon'_2 - \epsilon_2\epsilon'_1)
    \end{align}

This implies that the symplectic form generally can be written as:
\begin{align}
    \omega = & -\frac{c_1}{48\pi}\int_{S^1} \left[2S[f] d\epsilon \wedge d\epsilon'  - d\epsilon' \wedge d\epsilon''\right] -\frac{c_2}{24\pi}\int_{S^1} \Bigg[  S[f](d\epsilon \wedge d\sigma' - d\epsilon' \wedge d\sigma) \nonumber \\
    & + \left(\frac{\gamma'''}{f'} - \frac{3f''}{f'^2}\gamma'' + \frac{3f''^2}{f'^3}\gamma' - \frac{f'''}{f'^2}\gamma' \right) d\epsilon \wedge d\epsilon' +  d\epsilon'' \wedge d\sigma' \Bigg]+ \frac{h_j}{2\pi}\int_{S^1} f'^2 (d\epsilon \wedge d\epsilon') \nn\\
    &+  \frac{h_p}{2\pi}\int_{S^1} \left[f'^2(d\epsilon \wedge d\sigma' - d\epsilon' \wedge d\sigma) +f'\gamma' (d\epsilon \wedge d\epsilon') \right]
\end{align}
considering a one-form $d\epsilon \equiv \epsilon_n d\chi_n$ on the coadjoint space such that $d\chi_m\left(\xi_n\right) = \delta_{mn}$.
Here we can check that $\omega (\xi_1, \xi_2) = \omega_{12}$ written above. Using the group composition law, we can relate,
\begin{align}
    df = f' d\epsilon; \quad d\gamma = \gamma' d\epsilon + f' d\sigma 
\end{align}
Thus the symplectic form is given by,
\begin{align}
    \omega = & \frac{c_1}{48\pi}\int_{S^1}\left(\frac{1}{f'^2}df' \wedge df'' \right)
    - \frac{c_2}{24\pi}\int_{S^1} df' \wedge \left( - \frac{1}{f'^2} d\gamma'' + \frac{f''}{f'^3} d\gamma' + \frac{\gamma'}{f'^3}df'' \right)\nn\\
    & + \frac{h_j}{2\pi} \int_{S^1} (df\wedge df')+ \frac{h_p}{2\pi} \int_{S^1} \bigg(2df \wedge d\gamma' -\frac{\gamma'}{f'} df\wedge df' \bigg)\label{symp form}
\end{align}
Since we are interested in the one-loop quantization of the system, we consider infinitesimal fluctuations around the $f=\theta,\gamma=0$ solution of the theory,

\begin{align}
    f(\theta)=\theta + \sum_{\abs{n}\geq 1} \epsilon_n e^{i n \theta},\quad \gamma(\theta)=\sum_{\abs{n}\geq 1}\gamma_n e^{i n \theta}
\end{align}
Here we are considering ${\abs{n}\geq 1}$ since in the coadjoint orbit, the $u(1)\oplus u(1)$ transformations are quotiented and we only want to consider nontrivial BMS transformations. We will keep terms only till quadratic order in fluctuations. This gives us the infinitesimal version of \eqref{symp form}:
\begin{align}
    \omega = i\sum_{n\geq 1} n\bigg[\left(  \frac{c_1}{12}n^2+ 2h_j \right)d\epsilon_{-n}\wedge d\epsilon_n + \left( \frac{c_2}{12} n^2 +2h_p\right) (d\epsilon_{-n} \wedge d\gamma_n + d\gamma_{-n} \wedge d\epsilon_n )\bigg] \label{symp-inf}
\end{align}
Note that, for the Poincar\'{e}-invariant orbit representative \eqref{poinc inv rep}, the $n=\pm 1$ contributions vanish as expected.

The measure on the orbit $\mathcal{O}_{(h_j,h_p)}$ for constant $(h_j,h_p)$ with values \eqref{parameters-relation} is given by\footnote{The Pfaffian can be computed by rearranging the components of \eqref{symp-inf} into block matrices labeled by $n$.},
\begin{align}
    d\mu = \text{Pf}(\omega) d^2\epsilon d^2\gamma= \prod \limits_{n\ge 1} \bigg[\frac{n}{2}\left(\frac{c_2}{12} n^2 +2h_p \right)\bigg]^2 d^2\epsilon d^2\gamma. \label{measure}
\end{align}
Now We will evaluate the one-loop path integral using this measure.

\subsection{One-loop path integral}
The path integral corresponding to the theory \eqref{bms-PGB} is given as,
\begin{align}
    \mathcal{Z} (\beta,\Phi) = \int Df D\gamma \text{e}^{-I}
\end{align}
This describes the grand canonical partition function of a statistical system with inverse temperature $\beta$ and chemical potential $\Phi$. To evaluate the one-loop path integral, we expand the fields\footnote{The classical value of $\gamma$ being 0, we are denoting the fluctuations to be $\gamma(\theta)$ in the following computations.} around the classical solution as $f(\theta) = \theta + \epsilon(\theta)$ and expand the action \eqref{bms-PGB} to quadratic order. This gives,
\begin{align}
    I = I^{(0)} + I^{(2)}. 
\end{align}
Here $I^{(0)}$ is the onshell action \cite{Bhattacharjee:2023sfd},
\begin{align}
    I^{(0)} = -\frac{c_1 \pi}{6 \Phi} + \frac{c_2 \pi^2}{6}\frac{\beta}{\Phi^2}
\end{align}
$I^{(2)}$ is the quadratic action of fluctuations,
\begin{align}
    I^{(2)} = \int_{S^1} & \Big[\frac{\pi}{12\Phi^2}(c_2\beta - c_1\Phi)\epsilon'^{2}  - \frac{1}{48\pi}(c_2\beta + c_1\Phi)(3\epsilon''^2 + 2\epsilon'\epsilon''') \nonumber \\
    &+ \frac{c_2\Phi}{24\pi}\left( \frac{4\pi^2}{\Phi^2}\epsilon''-\epsilon^{(4)} \right)\gamma \Big]
\end{align}
Expanding the fields in terms of their Fourier modes, we get
\begin{align}
    & I^{(2)} = - \sum_{n\geq 1} \left[a_n\epsilon_{-n}\epsilon_n + b_n (\gamma_{-n}\epsilon_n + \gamma_n \epsilon_{-n})\right], \quad \text{with} \\
    & a_n =\frac{n^2}{24\Phi^2}(-4c_2\pi^2\beta + 4c_1\pi^2\Phi + c_2n^2\beta\Phi^2 + c_1 n^2\Phi^3), \quad b_n = \frac{c_2 n^2}{24 \Phi}(4\pi^2 + n^2\Phi^2)
\end{align}
Here again the sum excludes $n = 0$ as it corresponds to a gauge symmetry. The one-loop path integral is given by the following Gaussian integral with the measure \eqref{measure}:
\begin{align}
    \mathcal{Z} = \text{e}^{-I^{(0)}} \mathcal{Z}^{(2)}; \quad \mathcal{Z}^{(2)} = \int d\mu\ \text{e}^{-I^{(2)}}
\end{align}
Simplifying this, we get,
\begin{align}
    \mathcal{Z}^{(2)} = \prod \limits_{n\ge 1} \int  \bigg[\frac{n}{2}\left(\frac{c_2}{12} n^2 +\frac{c_2 \pi}{3 \Phi^2} \right)\bigg]^2 d^2\epsilon_n \, d^2\gamma_n \exp\bigg[ a_n\abs{\epsilon_n + \frac{b_n}{a_n}\gamma_n}^2 + \frac{b_n^2}{a_n}\abs{\gamma_n}^2\bigg].
\end{align}
We have used the value of $h_p$ \eqref{parameters-relation} in \eqref{measure}. Performing the $\epsilon$ Gaussian integrals keeping $\gamma$ fixed, and then finally performing the $\gamma$ integrals, we get
\begin{align}
        \mathcal{Z}^{(2)} &= \prod\limits_{n\ge 1} \frac{c^2 n^2}{(24)^2} \bigg(n^2 + \frac{4 \pi^2}{\Phi^2} \bigg)^2 \times \frac{(24\pi)^2}{c_2^2 n^4 \Phi^4} \bigg(n^2+\frac{4 \pi^2}{\Phi^2} \bigg)^{-2} \sim \prod\limits_{n\ge 1}\,\frac{1}{n^2 \Phi^2}.
\end{align}
This infinite product can be performed using zeta function regularization, as described in appendix \ref{zeta-reg}, we have:
\begin{align}
    \log \mathcal{Z}^{(2)} \sim \log\Phi, \quad \mathcal{Z}(\beta,\Phi) \sim \Phi \exp \bigg(-\frac{c_2 \pi^2}{6 \Phi^2} \beta + \frac{c_1 \pi^2}{ 6 \Phi}\bigg). 
\end{align}
Thus we see that the path integral receives logarithmic corrections from fluctuations at one-loop order. This is the most important result of this paper. For future convenience, we will express the one-loop partition function as,
\begin{align}
    \mathcal{Z}(\beta,\Phi) \sim z(\Phi) \exp \bigg(-\frac{c_2 \pi^2}{6 \Phi^2} \beta + \frac{c_1 \pi^2}{ 6 \Phi}\bigg). \label{1loop}
\end{align}
such that $z(\Phi) = \Phi$ gives us the partition function of the celestial dual theory. This will help us distinguish between nontrivial logarithmic contributions coming from one-loop effects and contributions coming from change of ensemble\footnote{In principle, the function $z$ could depend on $\beta$ as well. Then separating the two contributions would be complicated. However, evidently for us considering a dependence on $\Phi$ is enough.}. 

\section{Logarithmic corrections to entropy}\label{sec-ent}
Now, we compute the microcanonical entropy by taking the inverse Laplace transform of the partition function obtained above\footnote{We are using the usual sign in entropy as described in \cite{Barnich:2012aw, Barnich:2012xq}} in \eqref{1loop}, 
\begin{align}
    e^{-S(h_j,h_p)} &= \int d\beta\,d\Phi\,e^{\beta h_p+ \Phi h_j} \mathcal{Z} (\beta,\Phi)\nonumber\\
    &= \int d\Phi\, e^{\Phi h_j}\, z(\Phi) \, e^{\frac{c_1 \pi^2}{6 \Phi}}\, \int d\beta\, e^{\beta \big( h_p - \frac{c_2 \pi^2}{6 \Phi^2}\big)} 
\end{align}
Keeping in mind that for us, $z(\Phi) = \Phi$. The inverse Laplace transform in $\beta$ variable simply puts a delta function constraint on the $\Phi$ integral,
\begin{align}
e^{-S(h_j,h_p)} &= \int d\Phi\, e^{\Phi h_j}\, z(\Phi) \, e^{\frac{c_1 \pi^2}{6 \Phi}}\, \delta \bigg(h_p - \frac{c_2 \pi^2}{6\Phi^2} \bigg).
\end{align}
The delta function can be simplified as,
\begin{align}
    \delta \bigg(h_p - \frac{c_2 \pi^2}{6\Phi^2} \bigg) = \frac{3\Phi_0^3}{c_2\pi^2}[\delta(\Phi-\Phi_0) + \delta(\Phi+\Phi_0)]; \quad \Phi_0 = \sqrt{\frac{c_2 \pi^2}{6h_p}}
\end{align}
This essentially localizes the $\Phi$ integral to the values $\pm\Phi_0$. However, consistency with the semiclassical entropy of FSC fixes $\Phi = -\Phi_0$. Thus we pick this contribution only. Thus the density of states is given by,
\begin{align}
    e^{-S(h_j,h_p)} &= \frac{3\Phi_0^3}{c_2\pi^2} \exp \left(-\Phi_0 h_j-\frac{c_1 \pi^2}{6 \Phi_0}\right)  z(\Phi_0) \nonumber \\
    & \sim h_p^{-\frac{3}{2}}\exp \left[-\frac{\pi}{\sqrt{6}}\left(\sqrt{\frac{c_2}{h_p}}h_j + c_1\sqrt{\frac{h_p}{c_2}}\right)\right] z(\Phi_0(h_p))
\end{align}

The term inside the exponential is the negative of the Bekenstein-Hawking entropy \eqref{BH-ent}. Therefore, the entropy and its logarithmic corrections in the large charge regime are given as,
\begin{align}
    S(h_j,h_p) \sim S_{\text{BH}} + \frac{3}{2}\log h_p - \log z(\Phi_0(h_p)). \label{ent-z}
\end{align}
This shows us that the usual $\frac{3}{2}$ coefficient is solely an effect of change of ensemble whereas the nontrivial one-loop effects are encoded in $\log z$. In our case, $z(\Phi_0(h_p)) = \Phi_0 \sim h_p^{-1/2}$. Thus we get an additional factor of $\frac{1}{2}$ in the $\log h_p$ coefficient. The final expression from our celestial dual theory boils down to,
\begin{align}
    S(h_j,h_p) \sim S_{\text{BH}} + 2\log h_p. \label{ent}
\end{align}
This is the main result of this paper. The result is valid for both $c_1$ and $c_2$ being nonzero. Analyzing the scaling properties of the charges with horizon size $r_c$ of FSC solution in pure gravity (i.e. $c_1 = 0$), we get:
\begin{align}
    S_{\text{BH}} \sim r_c, \, h_p \sim r_c^{-2}, \, h_j \sim r_c^0
\end{align}
These can be easily extracted from the thermodynamic behavior and definition of charges of FSC, discussed in \cite{Barnich:2012aw}. This shows that in pure gravity, we have:
\begin{align}
    S(h_j,h_p) \sim S_{\text{BH}} - 4\log r_c. \label{log-hor}
\end{align}

\section{Discussions} \label{sec-concl}

In this paper, we have computed the path integral of the celestial dual theory \cite{Bhattacharjee:2023sfd} for (2+1)D asymptotically flat spacetimes. This dual theory captures the gravitational phase space near Flat Space Cosmological solutions in the large charge regime. We have found that the theory is one-loop exact due to presence of a Lagrange multiplier field. For the quantization, we have used coadjoint orbit techniques similar to Schwarzian theories \cite{Stanford:2017thb,Afshar:2019tvp,Afshar:2021qvi}. From the symplectic structure on the BMS$_3$ orbit with $U(1)\times U(1)$ stabilizer, we have first computed the path integral measure and then computed the one-loop path integral. Finally we have computed the logarithmic contributions to entropy using this path integral.

 While we get the log contributions from change of ensemble \cite{Bagchi:2013qva} naturally, our computation shows that additional log contributions are coming from the one-loop path integral itself. These nontrivial corrections were not considered in the past literature. 

The partition function around an FSC saddle was also computed in \cite{Barnich:2017jgw, Merbis:2019wgk} from 2D geometric actions on BMS$_3$ coadjoint orbits. These theories are related to the usual 2D Liouville-like duals. Finally, as a consistency check, we extract the log corrections from these results in the large charge regime and show that the coefficient matches our results. The grand canonical partition function of \cite{Merbis:2019wgk} has the form \eqref{1loop} at one-loop order, with 
\begin{align*}
    z(\Phi) = \prod_{n\geq 1} \frac{1}{\abs{1-q^n(\Phi)}^2}, \quad q (\Phi) \equiv e^{2\pi i \Phi}.
\end{align*}
Thus the log contribution in entropy has exactly the structure \eqref{ent-z} with the above form of $z$. Now let us understand the large charge behavior of the contribution coming from $z$. We note that in the large charge regime, the parameter $\Phi_0 \sim \sqrt{\frac{c_2}{h_p}} \ll 1$ such that,
\begin{align*}
    q (\Phi_0) \approx 1 + 2\pi i \Phi_0
\end{align*}
Thus the leading behavior of the function $z(\Phi_0)$ is,
\begin{align*}
    z(\Phi_0) \approx \prod_{n\geq 1} \frac{1}{\abs{1-( 1 + 2\pi i \Phi_0)^n}^2} \sim \prod_{n\geq 1} \frac{1}{\Phi_0^2 n^2}
\end{align*}
The higher-order terms are much suppressed. Again by using the product in appendix \ref{zeta-reg}, we get from \eqref{ent-z},
\begin{align*}
    & \log z(\Phi_0) \sim \log\Phi_0 \sim -\frac{1}{2}\log h_p \\
    & S(h_j,h_p) \sim S_{\text{BH}} + 2\log h_p.
\end{align*}
Thus the large charge behavior exactly matches our result \eqref{ent}.\\

Let us also compare our results with that of BTZ black holes which are asymptotically AdS$_3$ \cite{Govindarajan:2001ee, Carlip:2000nv}. The logarithmic correction to entropy for BTZ black holes is trivial and only arises due to a change of ensemble. From the bulk point of view this triviality was explained in \cite{Sen:2012dw} through Euclidean gravity formalism. The argument relies on the scaling properties of bulk charges with the horizon size and the presence of cosmological constant. Although the Flat space cosmologies can be understood as limits of BTZ solutions, these scaling properties change non-trivially and cannot be smoothly mapped into one another. It is important to find these novel corrections from alternative approaches e.g. from bulk fluctuations. The asymptotic symmetries near black hole/cosmological horizons determine their entropy \cite{Donnay:2015abr,Donnay:2016ejv,Lust:2017gez}. It would be interesting to understand whether these modes can give rise to the log corrections obtained from our method. This is an ongoing work.

\acknowledgments

We are grateful to Nabamita Banerjee for useful discussions and comments on the work. This project was supported in part by the Polish National Agency for Academic Exchange under the NAWA Chair programme. We would like to thank the people of India for their continuous support towards fundamental research.

\appendix

\section{Evaluation of infinite product}\label{zeta-reg}
In this section, we will regularize the following infinite product which has appeared in our computations,
\begin{align}
    Z = \prod_{n\geq 1}^{\infty} \lambda_n \label{Zprod}
\end{align}
For us, $\lambda_n$ is of the form $\frac{1}{\Lambda^2 n^2}$. For this, we apply the zeta function regularization trick and consider the following series in stead,
\begin{align}
    Y(s) = \sum_{n\geq 1}^{\infty} \lambda_n^s \label{Ysum}
\end{align}
Here $s$ is a complex number. The trick is to compute the series $Y(s)$ in an open domain of the complex plane and then analytically continue it near $s = 0$. Then we can express the regularized series $\log Z$ in terms of $Y(s)$,
\begin{align}
    Y'(s) = \sum_{n\geq 1}^{\infty} \lambda_n^s\log\lambda_n; \quad \log Z \stackrel{\text{reg.}}{=} Y'(0)
\end{align}
For our case i.e. $\lambda_n = \frac{1}{\Lambda^2 n^2}$,
\begin{align}
    Y(s) = \Lambda^{-2s}\zeta(2s), \quad Y'(s) = -2\Lambda^{-2s}\zeta(2s)\log\Lambda + 2 \Lambda^{-2s}\zeta'(2s)
\end{align}
This is analytic in the open set Re$(s)>\frac{1}{2}$. We can easily analytically continue this function near $s = 0$. Using $\zeta(0) = -\frac{1}{2}$, we get:
\begin{align}
    \log Z = Y'(0) \sim \log\Lambda.
\end{align}
We will use this logarithmic behavior in our main computations.

\bibliography{CelPF}
\end{document}